\documentclass[a4paper,11pt]{article}
\pdfoutput=1
\usepackage{graphicx}
\usepackage{amsmath}
\usepackage{amssymb}
\usepackage{latexsym}
\usepackage{color}
\usepackage{float}
\usepackage{cite}
\usepackage{makeidx}
\usepackage{colortbl}
\usepackage{multirow}

\newcommand{\bra}{\begin{array}}
\newcommand{\era}{\end{array}}
\newcommand{\beq}{\begin{equation}}
\newcommand{\eeq}{\end{equation}}
\newcommand{\bqr}{\begin{eqnarray}}
\newcommand{\eqr}{\end{eqnarray}}

\def\BC{\bb C}
\def\_\BC{\bbi C}

\def \lp {\left(}
\def \gp {\right)}

\def\no2 {{\textstyle{n\over 2}}}

\newcommand{\lam}{\lambda}
\newcommand{\si}{\sigma}

\newcommand{\De}{\Delta}

\newcommand{\be}{\beta}

\newcommand{\al}{\alpha}

\newcommand{\lb}{\label}
\newcommand{\nn}{\nonumber}

\begin{document}
\begin{titlepage}
\setcounter{page}{1}
\renewcommand{\thefootnote}{\fnsymbol{footnote}}

\begin{flushright}
\end{flushright}

\vspace{5mm}
\begin{center}

{\Large \bf {Goos-H\"anchen like Shifts in Graphene Double Barriers}}

\vspace{5mm}

{\bf Ahmed Jellal\footnote{\sf
ajellal@ictp.it -- a.jellal@ucd.ac.ma}}$^{a,b,c}$, {\bf Ilham Redouani}$^{c}$,
{\bf Youness Zahidi}$^{c}$ and {\bf Hocine Bahlouli}$^{a,d}$

\vspace{5mm}

{$^a$\em Saudi Center for Theoretical Physics, Dhahran, Saudi
Arabia}

{$^b$\em Physics Department, College of Science, King Faisal University,\\
PO Box 380, Alahsa 31982, Saudi Arabia}

{$^{c}$\em Theoretical Physics Group,  
Faculty of Sciences, Choua\"ib Doukkali University},\\
{\em PO Box 20, 24000 El Jadida, Morocco}

{$^d$\em Physics Department,  King Fahd University
of Petroleum $\&$ Minerals,\\
Dhahran 31261, Saudi Arabia}


\vspace{3cm}

\begin{abstract}

We study the Goos-H\"anchen like shifts for Dirac fermions in
graphene scattered by double barrier structures. After obtaining
the solution for the energy spectrum, we use the boundary
conditions to explicitly determine the Goos-H\"anchen like shifts
and the associated transmission probability. We analyze these two
quantities at resonances by studying their {main} characteristics
as a function of  the energy and electrostatic potential
parameters. {To check the validity of our computations we recover previous results
obtained for a single barrier under appropriate limits.}

\end{abstract}
\end{center}

\vspace{3cm}


\end{titlepage}


\section{Introduction}

Graphene remains among the most fascinating and
attractive subject in condensed matter physics~\cite{Novoselov}. This is
because of its exotic physical properties and the apparent
similarity of its mathematical model to the one describing
relativistic fermions in two dimensions. As a consequence of this
relativistic-like behavior, particle could tunnel through very high
barriers in contrast to the conventional tunneling of
non-relativistic particles, an effect known in relativistic field
theory as Klein Tunneling. Such effect has already been observed
experimentally~\cite{Stander} in graphene systems.

The Goos-H\"anchen (GH) effect \cite{Goos} is a phenomenon that
originated in classical optics in which a light beam reflecting
off a surface is spatially shifted as if it had briefly penetrated
the surface before bouncing back. The interface has to separate
different dielectric materials (such as glass or water), and
absorption or transmission should be small enough to allow a
substantial reflected beam to form \cite{Foster}. The size of the
GH effect is proportional to the derivative of the refection phase
with respect to the angle of incidence. In addition to shifting
beam position, the GH effect can manifest itself in alterations of
differential cross sections \cite{Tran} of laser mode dynamics
\cite{Dutriaux} and of mode spectra \cite{Chowdhurry}.

Various experimental and theoretical investigations classified the
transport properties among the most interesting features of Dirac
fermions in graphene. Among these transport properties we cite the
quantum version of the GH effect originating from the reflection
of particles from interfaces. The latter has been studied at a
single graphene interface~\cite{Beenakker,Zhao} and in
graphene-based electric and magnetic potential barriers
\cite{Sharma}. Subsequently, it was shown that the modulation of
Goos-H\"anchen like (GHL) shifts can be controlled by varying the
electrostatic potential and induced gap~\cite{Chen4} and can even
be enhanced by transmission resonances. Very recently, the giant
GHL shifts for electron beams tunneling through graphene double
barrier structures~\cite{Song} was also investigated and it was
found that the shifts exhibits a sharp peaks inside the
transmission gap.

{There are various ways for creating barrier structures in
graphene~\cite{Katsnelsonn, Sevinçli}, for instance it can be done
by applying a gate voltage, cutting the graphene sheet into finite
width to create a nanoribbons, using doping or through the
creation of a magnetic barrier.}
{Different experimental methods are available to open a gap in graphene systems.
One of them is through an
inversion symmetry breaking of the sublattice due to the
fact that the densities of the particles associated with the
on-site energy for A and B sublattice are different \cite{zhou}.
As demonstrated in the experiment, the maximum energy
gap could be $260 meV$ due to the sublattice symmetry
breaking \cite{zhou}. Therefore, the periodic dependence
of GHL shifts on the induced gap also provides an efficient
way to modulate the lateral shifts in a fixed gapped
graphene barrier, which is useful for the manipulation of
electron beam propagation in graphene \cite{wang}.}

Based on previous investigations of Dirac fermions and in
particular our recent work \cite{Alhaidari,Choubabi,Mekkaoui},
where we developed our approach to deal with graphene double
barrier structures, in the present work we analyze the
GHL shifts by considering Dirac fermions in the presence of an
electrostatic potential placed between two regions composing the
graphene sheet.
For general purposes, we consider the potential
configuration depicted in Figure 1 rather than that used
in~\cite{Song}. By requiring the continuity of the wave functions
at interfaces, we show that it is possible to determine the GHL
shifts as well as the transmission probability in terms of the
incident angle. By focussing on resonances, we conclude that these
two quantities become large and their shapes exhibit different peaks.
We emphasis the difference between our results and those obtained
in~\cite{Song} by studying double barriers. Considering an
appropriate limiting case, we derive interesting results regarding
single barriers. Comments and discussions will be provided in the
main text to support the relevance of our present potential
configuration.

The paper is organized as follows. In section 2, we formulate our
model by setting the Hamiltonian system describing particles
scattered by double barrier structures. Considering the five
potential regions one at a time, we obtain the spinor solution corresponding to
each region in terms of different scattering potential parameters. Using
boundary conditions we are required to split the energy into three
domains in order to calculate the transmission probability in section 3 and
GHL shift in section 4 in each domain.  In each situation, we
analyze the GHL shifts and transmission at resonances that
characterize each region.  In section 5, we discuss the importance
of our numerical results and present different supporting plots. We also
study some limiting cases, particular interest is the situation
where our problem reduces to the single barrier problem. Finally, we
conclude our work in section 6 and emphasize our main results.

\section{Theoretical model}

\begin{figure}[h]
  \centering
  \includegraphics[width=12cm, height=5cm ]{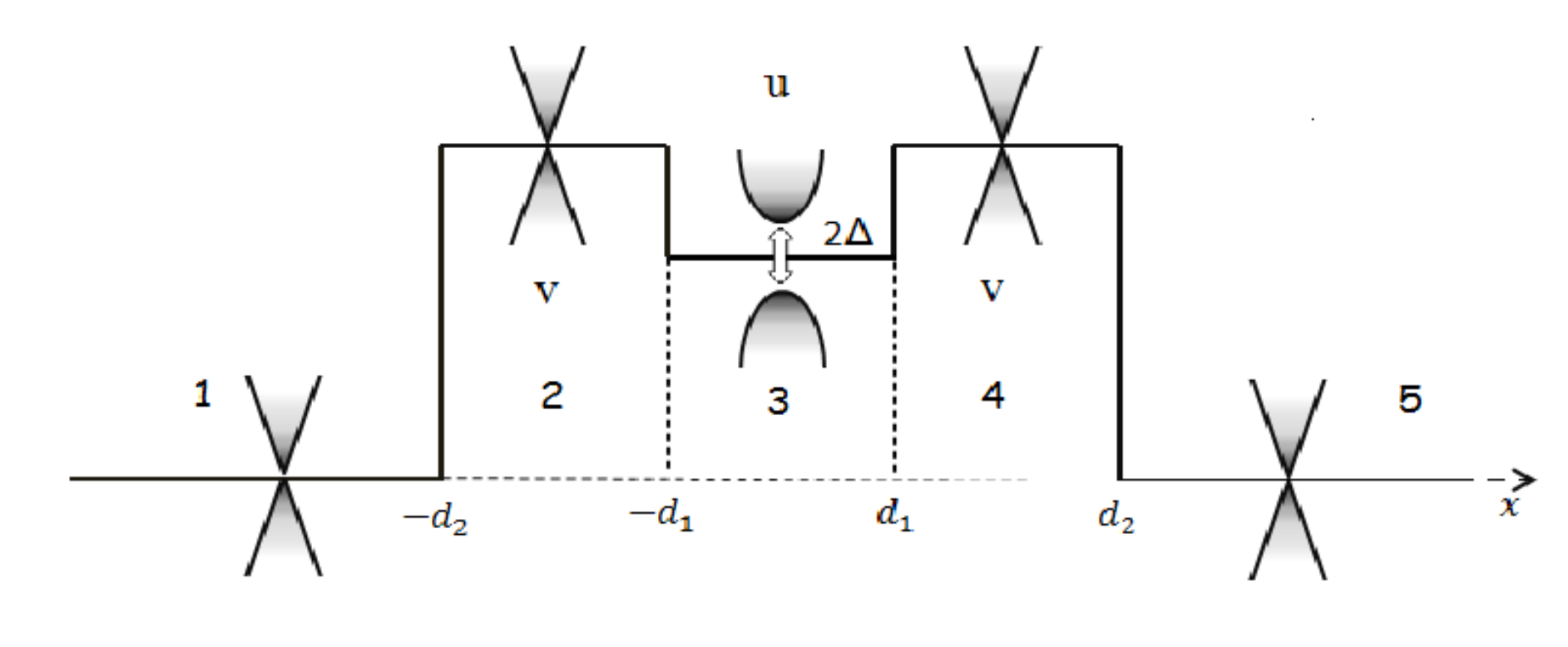}\\
  \caption{\sf Schematic diagram for the monolayer graphene double barrier.}\label{fig.1}
\end{figure}

We consider Dirac fermions in graphene scattered by an electrostatic double barrier potential.
We can write the Hamiltonian describing our particle in the central region 3 as follows
\beq\lb{3}
H_{\sf3}= v_{F} \vec\si \cdot \vec p + V_3(x)\mathbb{I}_{2}+ \De \si_z
\eeq
where ${v_{F}\approx 10^6 m/s}$  is the Fermi velocity, ${\sigma =(\sigma_{x},\sigma_{y})}$  are the
Pauli matrices and $\vec{p}=-i\hbar\vec{\nabla}$. The parameter $\Delta = m v_{F}^2$ is the energy gap owing to the sublattice symmetry breaking or it
can be seen as the energy gap $\Delta = \Delta_{so}$ originating from spin-orbit interaction. Elsewhere the system is described by the Hamiltonian
\begin{equation}\label{eq 1}
 H_{j}=v_{F} \vec{\sigma}\cdot\vec{p}+V_j(x) \mathbb{I}_{2}.
\end{equation}
In order to study the scattering of Dirac fermions in graphene
by the above double barrier structure we first choose the following potential configuration for
the double barrier potential
\begin{equation} \label{eq 2}
V_j(x)=\left\{\begin{array}{lll} {v} & \mbox{if} & {d_{1}<\lvert x \rvert<d_{2}} \\
{u} & \mbox{if} & {\lvert x \rvert < d_{1} } \\ {0} & {} &
\mbox{otherwise} \end{array}\right.
\end{equation}
where $j$ labels the five regions indicated schematically in Figure \ref{fig.1},
 which shows the space configuration of the potential profile

Solving the eigenvalue equation to obtain the upper and lower
components of the eigenspinor in the incident and reflection regions
{\sf 1} ($x < - d_{2}$)
\begin{equation}\label{eq3}
    \Phi_{\sf 1}=  \left(
            \begin{array}{c}
              {1} \\
              {z_{1}} \\
            \end{array}
          \right) e^{i(k_{1}x+k_{y}y)} + r\left(
            \begin{array}{c}
              {1} \\
              {-z_{1}^{-1}} \\
            \end{array}
          \right) e^{i(-k_{1}x+k_{y}y)}.
\end{equation}
In region {\sf 2} ($-d_{2}<x<-d_{1}$), we obtain the solution
\begin{equation}\label{eq4}
 \Phi_{\sf 2}= a \left(
            \begin{array}{c}
              {1} \\
              {z_{2}} \\
            \end{array}
          \right) e^{i(k_{2}x+k_{y}y)} +b \left(
            \begin{array}{c}
              {1} \\
              {-z_{2}^{-1}} \\
            \end{array}
          \right) e^{i(-k_{2}x+k_{y}y)}.
\end{equation}
In region {\sf 4} ($d_{1}<x<d_{2}$), we have the eigenspinor
 \begin{equation}\label{eq5}
 \Phi_{\sf 4}= e \left(
            \begin{array}{c}
              {1} \\
              {z_{2}} \\
            \end{array}
          \right) e^{i(k_{2}x+k_{y}y)} +f \left(
            \begin{array}{c}
              {1} \\
              {-z_{2}^{-1}} \\
            \end{array}
          \right) e^{i(-k_{2}x+k_{y}y)}
\end{equation}
and finally the eigenspinor in region {\sf 5} ($x > d_{2}$) can be expressed as
\begin{equation}\label{eq6}
 \Phi_{\sf 5}= t \left(
            \begin{array}{c}
              {1} \\
              {z_{1}} \\
            \end{array}
          \right) e^{i(k_{1}x+k_{y}y)}.
\end{equation}
We have introduced the perpendicular $k_{j} =
k_{{F}_j}\cos\theta_{j}$ and parallel
$k_{y}=k_{{F}_j}\sin\theta_{j}$ components of the wave vector as
well as the  parameter \beq z_{j}
=s_{j}\frac{k_{j}+ik_{y}}{\sqrt{k_{j}^{2}+k_{y}^{2}}}=s_{j}e^{i\theta_{j}}
\eeq
where the
sign function is defined by
$s_{j}={\mbox{sign}}{\left(E-V_j\right)}$ with the phase
$\theta_{j}=\arctan\left(k_{y}/k_{j}\right)$ and $k_{F_j}=
\frac{2\pi}{\lam_j}$ is the Fermi wave vector. The corresponding
dispersion relation in these four regions is given by \beq
E-V_j=s_j\sqrt{k_j^2 +k_y^2}=s_j\hbar v_F k_{F_j}. \eeq

Now solving the eigenvalue equation for the Hamiltonian \eqref{3} describing region 3,
we find the following eigenspinor
\begin{equation} \label{eq 7}
 \Phi_{\sf 3}= c_1 \left(
            \begin{array}{c}
              {\alpha} \\
              {\beta z_{3}} \\
            \end{array}
          \right) e^{i(k_{3}x+k_{y}y)} +c_2 \left(
            \begin{array}{c}
              {\alpha} \\
              {-\beta z_{3}^{-1}} \\
            \end{array}
          \right) e^{i(-k_{3}x+k_{y}y)}
\end{equation}
with the parameters $\alpha$ and $\beta$ are given by
\begin{equation} \label{eq 8}
       {\alpha=\lp{1+\frac{s_{3} \Delta}{\sqrt{\Delta^{2}+{\hbar}^{2}{v_{F}}^{2} {k_{F_3}}^{2}}}}\gp}^{1/2}, \qquad
       \beta=\lp{1-\frac{s_{3} \Delta}{\sqrt{\Delta^{2}+{\hbar}^{2}{v_{F}}^{2} {k_{F_3}}^{2}}}}\gp^{1/2}
 \end{equation}
the complex number is defined by
\beq
z_{3}=s_{3}e^{i\theta_{3}}, \qquad \theta_{3}=\arctan\left(k_{y}/k_{3}\right)
\eeq
with the sign function
 $s_{3}=\mbox{sign}(E-u)$.
The Fermi wave vector being defined by
\beq
k_{F_3}=\frac{1} {\hbar v_{F}} \sqrt{(E-u)^{2}-\Delta^{2}}
\eeq
and $k_3$ is given by the relation
\beq
 k_{3}=\sqrt{{k_{
F_3}^{2}-{k_{y}}^{2}}}.
\eeq
As usual the coefficients $(c_1,c_2)$ together with  $(a,b,e,f,r,t)$ can be determined using
the boundary conditions, continuity of the eigenspinors at each interface. Next we will use
the above solutions to compute the transmission coefficient and associated phase shift and build
a bridge between quantum optics and Dirac fermions in graphene.

 \section{Transmission amplitude and phase shift}

Before determining explicitly the transmission coefficient and the
phase shift, let us discuss the main features of the
solutions obtained above. We characterize our waves by introducing
a critical angle $\phi_{c}$, defined by
\beq \phi_{c} = \arcsin
\sqrt{\frac{k_{F_2}}{k_{F_1}}} = \arcsin \frac{\sqrt{(E-v)^2 }}{E}
\eeq
which corresponds to total internal reflection. We notice
that when the incident angle is less than  $\phi_{c}$, i.e
$\theta_{1}<\phi_{c}$, the modes become oscillating guided modes. While in
the case when the incident angle is more than  $\phi_{c}$, i.e
$\theta_{1}>\phi_{c}$, we have decaying or
evanescent wave modes, this amounts to replace $k_{3}$ by $i \kappa$
with \beq
 \kappa=\sqrt{{k_{y}}^{2}-{k_{F_3}}^{2}}.
\eeq

In the forthcoming analysis, we will be interested in studying the situation $ \theta_{1}<\phi_{c}$
and investigate the GHL shifts for particles scattered by the potential profile \eqref{eq 2}.
This will necessitate the evaluation of the transmission coefficient and the phase shift.
Matching the wave functions at the boundaries $(-d,-d/2,d/2,d)$ as required by the first order nature
of the Dirac equation we end up with the following set of equations
\bqr \label{eq11}
&&e^{-ik_{1}d}+re^{ik_{1}d}= ae^{-ik_{2}d}+be^{ik_{2}d}\\ 
&&ae^{-ik_{2}d/2}+be^{ik_{2}d/2}=c_1\alpha e^{-ik_{3}d/2}+c_2\beta e^{ik_{3}d/2}\\  
&&c_1\alpha e^{ik_{3}d/2}+c_2\beta e^{-ik_{3}d/2}=ee^{ik_{2}d/2}+fe^{-ik_{2}d/2}\\
&&ee^{ik_{2}d}+fe^{-ik_{2}d}=te^{ik_{1}d}. 
\eqr
After a lengthy but straightforward algebra, we can show that the transmission coefficient can be written
in terms of the phase shift $\varphi$ as indicated below
 \beq
t=\frac{e^{i\varphi}}{f_{0}}
\eeq
and we have 
 \bqr \label{eq 12}
     f_{0} e^{i \varphi} &=& \chi_1\cos\left[(k_2-k_3)d\right]+(1-\chi_1)
             \cos\left[(k_2+k_3)d\right]\\
             && +  i\left\{\left(-\chi_1
             \sin\left[(k_2-k_3)d\right]+(\chi_1-1)\sin\left[(k_2+k_3)d\right]\right)\chi_2
             -\chi_3\chi_4\sin\left(k_{3}d\right)\right\}\nonumber
\eqr
where the $\chi$'s are given by
 \bqr \label{eq 13}
          \chi_1 &=& \frac{1}{2} \left(1+\tan{\theta_2}\tan{\theta_3}- \frac{s_2s_3}{\al\be} 
           \sec{\theta_2}\sec{\theta_3}\right)\nn\\
          \chi_2 &=& \tan{\theta_1}\tan{\theta_2}-s_1s_2\sec{\theta_1}\sec{\theta_2}\nn\\
          \chi_3 &=& s_1\sec{\theta_1}\tan{\theta_2}-s_2\sec{\theta_2}\tan{\theta_1}\\
          \chi_4 & =&-s_2\sec{\theta_2}\tan{\theta_3}+ \frac{s_3}{\al\be} 
          \sec{\theta_3}\tan{\theta_2}.\nn
     \eqr
Using (\ref{eq 12}) and (\ref{eq 13}), we can show that the phase shift can be expressed explicitly as follows
 \begin{equation} \label{eq 14}
         \varphi=\arctan{\left(-\frac{\left(\chi_1 \sin\left[(k_2-k_3)d\right]+(1-\chi_1)\sin\left[(k_2+k_3)d\right]\right)\chi_2
         +\chi_3\chi_4\sin\left(k_{3}d\right)}{
         \chi_1\cos\left[(k_2-k_3)d\right]+(1-\chi_1)
         \cos\left[(k_2+k_3)d\right]}\right)}.
 \end{equation}
We will see how this phase will be used to investigate the GHL shifts for
Dirac fermions scattered by double barriers as depicted in Figure 1. The computation of the GHL shift will be performed in the
next section where we will also emphasis the main difference between our results and those obtained by Song \cite{Song}.

\section{GHL shifts through double barriers}

We begin our study of the GHL shift by considering an incident, reflected and
transmitted beams around some transverse wave vector $k_y = k_{y_0}$ corresponding to the central incidence angle $\theta_{1_0}$, denoted by the subscript $0$. These can be expressed in integral forms as follows
\begin{eqnarray}
   \Psi_{i}(x,y) &=& \int_{-\infty}^{+\infty}dk_y\ g(k_y-k_{y_0})\ e^{i(k_1(k_y)x+k_yy)}\left(
            \begin{array}{c}
              {1} \\
              {z_{1}} \\
            \end{array}
          \right)\label{inci}\\
\Psi_{r}(x,y) &=& \int_{-\infty}^{+\infty}dk_y\ r(k_y)\ g(k_y-k_{y_0})\ e^{i(-k_1(k_y)x+k_yy)}\left(
            \begin{array}{c}
              {1} \\
              {-z_{1}^{-1}} \\
            \end{array}
          \right)\label{refl}\\
   \Psi_{t}(x,y) &=& \int_{-\infty}^{+\infty}dk_y\ t(k_y)\ g(k_y-k_{y_0})\ e^{i(k_1(k_y)x+k_yy)}\left(
            \begin{array}{c}
              {1} \\
              {z_{1}} \\
            \end{array}
          \right)\label{trans}
\end{eqnarray}
where each spinor plane wave is a solution of \eqref{eq 1} and $g(k_y-k_{y_0})$ is the angular spectral distribution, which can be assumed of Gaussian
shape $w_ye^{-w_{y}^2(k_y-k_{y_0})^2}$ with $w_y$ being the half beam width at waist \cite{Beenakker}. The reflection $r(k_y)$ and transmission $t(k_y)$ coefficients will be calculated through the use of boundary conditions.

In order to calculate the GHL shifts of the transmitted beam through the graphene double
barriers, we adopt the definition \cite{Chen1,Chen2}
 \begin{equation} \label{eq 15}
        s_{t}=- \frac{\partial \varphi}{\partial k_{y_0}}.
 \end{equation}
We emphasis that our calculation will be done for different values
of the signature $s_j$ in the three regions of interest but at first we
consider a zero gap, i.e. $\Delta = 0$. Clearly, the value of
$s_j$ can be obtained from the relation
\beq s_j= \frac{ E-V_j}
{\sqrt{k_{j_0}^{2} + k_{y_0}^{2}}}
\eeq
which  tells us that
$s_j = 1$ on the particle-like branch,
$E > V_j$,
whereas $s_j = -1$ on the hole-like branch,  $E < V_j$. Therefore, to discuss the relevance of our results, we consider
different cases as summarized below in  Table \ref{tabl1}  \\

\begin{table}[h]
 \begin{center}
\begin{tabular}{|c|l|l|c|c|}
\hline
$ \  \mbox{Region 1} $\ \ & $\  \mbox{Region 2}$\ \ & $ \  \mbox{Region 3}$\ \  & \ $ E(u<v)$ \ \ &\ $  E(u>v)$ \ \ \\
\hline \hline
\multirow{4}*{$s_1=1$}&\multirow{2}*{$s_2=-1$} & $s_3=-1$ & $E<u<v$ &$E<v<u$\\
\cline{3-5}
$$ & $$ & $s_3=1$ & $u<E<v$ &\cellcolor[gray]{0.5}\\
\cline{2-5}
$$ & \multirow{2}*{$s_2=1$} & $s_3=-1$ & \cellcolor[gray]{0.5}&$v<E<u$ \\
\cline{3-5}
$$ & $ $ & $s_3=1$ & $u<v<E$& $v<u<E$\\
\hline
\end{tabular}
\end{center}
\caption{ \sf Different signature cases $s_j$ and their associated energy intervals.}\label{tabl1}
\end{table}
Now according to the above table, one has to consider three different cases in
order to determine the GHL shifts in our system. These cases are characterized by their energy regions as follows: $E<u<v$,
$u<E<v$  and finally $u<v<E$. Note that, we are focussing here only on $u < v$
while the case $u > v$ will be considered later in section 6.

$\blacktriangleright$ For  case 1: $E<u<v$, we use \eqref{eq 15} to obtain the following
GHL shifts
 \begin{equation}\label{eq 17}
        s_{t}=-\frac{d^2} {f_0^2} \tan{\theta_{1_0}}\left(-A_1B_1+C_1D_1\right)
 \end{equation}
where different quantities are defined by
 \begin{equation} \label{eq 18}
    \begin{array}{llll}
           A_1=\frac
           {1+ \mu_{-}^2} 
           {k_{2_0}d} \cos\left(k_{3_0}d\right) 
           \sin\left(k_{2_0}d\right)+
          \frac{1+ \nu_{-}^2} {k_{3_0}d} 
            \cos\left(k_{2_0}d\right) {\sin\left(k_{3_0}d\right)}+
           \frac{2- \mu_{-}^2- \nu_{-}^2} {k_{2_0}k_{3_0}d^2}  
           {\sin\left(k_{2_0}d\right)}
           {\sin\left(k_{3_0}d\right)}\\
           B_1=\frac{k_{{12}}^+ k_{{23}}^-k_{y_0}^2}{k_{2_0}^2k_{3_0}}\sin\left(k_{3_0}d\right)+
           \frac{k_{0_+}^{2}}{k_{2_0}}\left(\cos\left(k_{3_0}d\right)\sin\left(k_{2_0}d\right)+
           \frac{k_{3_0}\mu_{-}^2}{k_{2_0}}\sin\left(k_{3_0} d\right)\cos\left(k_{2_0}d\right)\right)\\
           C_1=\cos\left(k_{2_0}d\right)\cos\left(k_{3_0}d\right)-\frac{k_{3_0} \mu_{-}^2}{k_{2_0}}
           \sin\left(k_{2_0}d\right)\sin\left(k_{3_0}d\right)\\
           ¨
           {D_1=\frac{k_{0_+}^{2}}{k_{2_0}}
           \left(- \frac{1+\mu_{-}^2} {k_{2_0}d}
          {\cos\left(k_{2_0}d\right)\cos\left(k_{3_0}d\right)}+\frac{1+\nu_{-}^2} {k_{3_0}d}
           {\sin\left(k_{2_0}d\right)\sin\left(k_{3_0}d\right)}-
          \frac{2-\mu_{-}^2- \nu_{-}^2} {k_{2_0}k_{3_0}d^2}
           {\cos\left(k_{2_0}d\right)\sin\left(k_{3_0}d\right)}\right)}\\
           {\qquad +
           \frac{k_{12}^+ k_{23}^-}{k_{2_0}^2 k_{3_0}d^2}\left({2\sin{k_{3_0}d}}+
           k_{y_0}^2\left(\frac{1}{k_{1_0}^2}+\frac{2}{k_{2_0}^2}+\frac{1}{k_{3_0}^2}\right)\sin{\left(k_{3_0}d\right)}-
           \frac{k_{y_0}^2d}{k_{3_0}}\cos\left(k_{3_0}d\right)\right)}\\
           {\qquad + \frac{1} {k_{2_0}d^2} \left(2+\frac{k_{0_+}^{2}}{k_{2_0}^2}+\frac{k_{0_+}^{2}}{k_{1_0}^2}\right)
           \left(
           {\cos\left(k_{3_0}d\right)\sin\left(k_{2_0}d\right)+\frac{k_{3_0} \mu_{-}^2}{k_{2_0}}\cos\left(k_{2_0}d\right)
           \sin\left(k_{2_0}d\right)}
            \right)}\nonumber
       \end{array}
 \end{equation}
and we have set
\bqr
&& \mu_{\pm} =\sqrt{k_{F_2}k_{F_3}\pm k_{y_0}^2}/k_{3_0}, \qquad   \nu_{\pm} =\sqrt{k_{F_2}k_{F_3}\pm k_{y_0}^2}/k_{2_0} \\
 && k_{{12}}^\pm= k_{F_1}\pm k_{F_2}, k_{{23}}^\pm= k_{F_2}\pm k_{F_3}, \qquad  k_{0_\pm}=\sqrt{k_{F_1}k_{F_2}\pm k_{y_0}^2}.
 \eqr
The corresponding transmission probability reads
 \begin{eqnarray} \label{eq19}
     T 
       &=&\left[\left(\cos\left(k_{2_0}d\right)\cos\left(k_{3_0}d\right)-
       \frac{k_{3_0} \mu_{-}^2}{k_{2_0}}\sin\left(k_{2_0}d\right)\sin\left(k_{3_0}d\right)\right)^2\right.\\
       &&+\left.\left(\frac{k_{12}^+ k_{23}^-k_{y_0}^2}{k_{1_0}k_{2_0}^2k_{3_0}}\sin\left(k_{3_0}d\right)
       +\frac{k_{0_+}^{2}}{k_{1_0}k_{2_0}}\left(\sin\left(k_{2_0}d\right)\cos\left(k_{3_0}d\right)+
       \frac{k_{3_0} \mu_{-}^2}{k_{2_0}}\sin\left(k_3d\right)\cos\left(k_{2_0}d\right)\right)\right)^2\right]^{-1}\nn.
 \end{eqnarray}
It is clearly seen that to obtain full transmission one should impose resonance conditions, which are given by
 \beq
 k_{2_0}d = N_1\pi,  \qquad
 k_{3_0}d=N_2\pi
 \eeq
 with  $N_1, N_2 = 0,\pm 1,\pm 2, \cdots$. Under these circumstances, 
\eqref{eq 17} can be reduced  to the simple form
 \begin{equation} \label{eq 20}
      s_{t}|_{k_{2_0}d=N_1\pi,{k_{3_0}d=N_2\pi}}=d\tan\theta_{1_0}\left[\frac{k_{0_+}^{2}}{k_{2_0}^2}\left(1+\mu_{-}^2\right)+
      (-1)^{N_1}\frac{k_{12}^+k_{23}^-k_{y_0}^2}{k_{2_0}^2k_{3_0}^2}\right]
 \end{equation}
which corresponds to the maximum absolute value of the GHL shift.

$\blacktriangleright$
Now we consider the case 2:  $u<E<v$, which gives
 \begin{equation}\label{eq 26}
        s_{t}=-\frac{d^2}{f_0^2}\tan{\theta_{1_0}}\left(-A_2B_2+C_2D_2\right)
 \end{equation}
with
\begin{equation} \label{eq 27}
    \begin{array}{llll}
           A_2=\frac{1-\mu_{+}^2}{k_{2_0}d}\cos\left(k_{3_0}d\right)\sin\left(k_{2_0}d\right)+
           \frac{1-\nu_{+}^2}{k_{3_0}d}\cos\left(k_{2_0}d\right)\sin\left(k_{3_0}d\right)+
           \frac{2+\mu_{+}^2 +\nu_{+}^2}{k_{2_0}k_{3_0}d^2}\sin\left(k_{2_0}d\right)
           \sin\left(k_{3_0}d\right)\\
           B_2=\frac{k_{12}^{+}k_{23}^{+}k_{y_0}^2}{k_{2_0}^2k_{3_0}}\sin\left(k_{3_0}d\right)+
           \frac{k_{0_+}^{2}}{k_{2_0}}\left(\cos\left(k_{3_0}d\right)\sin\left(k_{2_0}d\right)-
           \frac{k_{3_0}\mu_{+}^2}{k_{2_0}}\sin\left(k_{3_0} d\right)\cos\left(k_{2_0}d\right)\right)\\
           C_2=\cos\left(k_{2_0}d\right)\cos\left(k_{3_0}d\right)+\frac{k_{3_0}\mu_{+}^2}{k_{2_0}}\sin\left(k_{2_0}d\right)\sin\left(k_{3_0}d\right)\\
           {D_2=\frac{k_{0_+}^{2}}{k_{2_0}}\left(-\frac{1-\mu_{+}^2}{k_{2_0}d}\cos\left(k_{2_0}d\right)\cos\left(k_{3_0}d\right)+
           \frac{1-\nu_{+}^2}{k_{3_0}d}\sin\left(k_{2_0}d\right)\sin\left(k_{3_0}d\right)-
           \frac{2+\mu_{+}^2+
          \nu_{+}^2}{k_{2_0}k_{3_0}d^2}\cos\left(k_{2_0}d\right)\sin\left(k_{3_0}d\right)\right)}\\{\qquad+
           \frac{k_{12}^{+}k_{23}^{+}}{k_{2_0}^2k_{3_0}d^2}\left(2\sin{k_{3_0}d}+
           k_{y_0}^2\left(\frac{1}{k_{1_0}^2}+\frac{2}{k_{2_0}^2}+\frac{1}{k_{3_0}^2}\right)\sin{\left(k_{3_0}d\right)}-
           \frac{k_{y_0}^2d}{k_{3_0}}\cos\left(k_{3_0}d\right)\right)}\\{\qquad+\frac{1}{k_{2_0}d^2}
            \left(2+\frac{k_{0_+}^{2}}{k_{2_0}^2}+\frac{k_{0_+}^{2}}{k_{1_0}^2}\right)\left(
           \cos\left(k_{3_0}d\right)\sin\left(k_{2_0}d\right)-\frac{k_{3_0}\mu_{+}^2}{k_{2_0}}\cos\left(k_{2_0}d\right)
           \sin\left(k_{3_0}d\right)\right)}\nonumber
       \end{array}
 \end{equation}
while the transmission probability is given by
 \begin{equation} \label{eq 28}
    \begin{array}{ll}
          T 
          =\left[\left(\cos\left(k_{2_0}d\right)\cos\left(k_{3_0}d\right)+
         \frac{k_{3_0}\mu_{+}^2}{k_{2_0}}\sin\left(k_{2_0}d\right)\sin\left(k_{3_0}d\right)\right)^2
          +\left(\frac{k_{12}^{+}k_{23}^{+}k_{y_0}^2}{k_{1_0}k_{2_0}^2k_{3_0}}\sin\left(k_{3_0}d\right)
           \right.\right.\\
          \left. \left. \qquad +\frac{k_{0_+}^{2}}{k_{1_0}k_{2_0}}\left(\sin\left(k_{2_0}d\right)\cos\left(k_{3_0}d\right)-
          \frac{k_{3_0}\mu_{+}^2}{k_{2_0}}\sin\left(k_{3_0}d\right)\cos\left(k_{2_0}d\right)\right)\right)^2\right]^{-1}
    \end{array}
  \end{equation}
The GHL shifts at resonances are given by
 \begin{equation} \label{eq 29}
       s_{t}|_{k_{2_0}d=N_1\pi,{k_{3_0}d
       =N_2\pi}}=d\tan\theta_{1_0}\left(\frac{k_{0_+}^{2}}{k_{2_0}^2}\left(1-\mu_+^2\right)+
       (-1)^{N_1}\frac{k_{12}^{+}k_{23}^{+}k_{y_0}^2}{k_{2_0}^2k_{3_0}^2}\right).
 \end{equation}


$\blacktriangleright$ For the case 3: $u<v<E$, the corresponding GHL shifts are found to be 
\begin{equation}\label{eq 34}
      s_{t}=-\frac{d^2}{f_0^2}\tan\theta_{1_0}\left(-A_3B_3+C_3D_3\right)
      \end{equation}
where various parameters are defined by
\begin{equation} \label{eq 35}
\begin{array}{llll}
       A_3=\frac{1+\mu_-^2}{k_{2_0}d}\cos\left(k_{3_0}d\right)\sin\left(k_{2_0}d\right)+
       \frac{1+\nu_-^2}{k_{3_0}d} \cos\left(k_{2_0}d\right)\sin\left(k_{3_0}d\right)+
        \frac{2-\mu_-^2-\nu_-^2}{k_{2_0}k_{3_0}d^2}\sin\left(k_{2_0}d\right)\sin\left(k_{3_0}d\right)\\
        {B_3=-\frac{k_{12}^-k_{23}^-k_{y_0}^2}{k_{2_0}^2k_{3_0}}\sin\left(k_{3_0}d\right)-
        \frac{k_{0_-}^2}{k_{2_0}}\left(\cos\left(k_{3_0}d\right)\sin\left(k_{2_0}d\right)+
        \frac{k_{3_0}\mu_-^2}{k_{2_0}}\sin\left(k_{3_0}d\right)\cos\left(k_{2_0}d\right)\right)}\\
        {C_3=\cos\left(k_{2_0}d\right)\cos\left(k_{3_0}d\right)- \frac{k_{3_0}\mu_-^2}{k_{2_0}}\sin\left(k_{2_0}d\right)\sin\left(k_{3_0}d\right)}\\
        {D_3=\frac{k_{0_-}^2}{k_{2_0}}\left(\frac{1+\mu_-^2}{k_{2_0}d}\cos\left(k_{2_0}d\right)\cos\left(k_{3_0}d\right)
       -\frac{1+\nu_-^2}{k_{3_0}d}\sin\left(k_{2_0}d\right)\sin\left(k_{3_0}d\right)+
        \frac{2-\mu_-^2-\nu_-^2}{k_{2_0}k_{3_0}d^2}\cos\left(k_{2_0}d\right)\sin\left(k_{3_0}d\right)\right)}\\{\qquad-
        \frac{k_{12}^-k_{23}^-}{k_{2_0}^2k_{3_0}d^2}\left(2\sin\left(k_{3_0}d\right)+
       k_{y_0}^2\left(\frac{1}{k_{1_0}^2}+\frac{2}{k_{2_0}^2}+\frac{1}{k_{3_0}^2}\right)\sin\left(k_{3_0}d\right)-
       \frac{k_{y_0}^2d}{k_{3_0}}\cos\left(k_{3_0}d\right)\right)}\\
        {\qquad+\frac{1}{k_{2_0}d^2}\left(2-\frac{k_{0_-}^2}{k_{2_0}^2}-\frac{k_{0_-}^2}{k_{1_0}^2}\right)
        \left(\cos\left(k_{3_0}d\right)\sin\left(k_{2_0}d\right)+
        \frac{k_{3_0}\mu_-^2}{k_{2_0}}\cos\left(k_{2_0}d\right)\sin\left(k_{3_0}d\right)\right)
        }.\nonumber
    \end{array}
\end{equation}
The transmission probability reduces to
\begin{equation} \label{eq 36}
      \begin{array}{ll}
            T 
             =\left[\left(\cos\left(k_{2_0}d\right)\cos\left(k_{3_0}d\right)-
        \frac{k_{3_0}\mu_-^2}{k_{2_0}}\sin\left(k_{2_0}d\right)\sin\left(k_{3_0}d\right)\right)^2
        +\left(-\frac{k_{12}^-k_{23}^-k_{y_0}^2}{k_{1_0}k_{2_0}^2k_{3_0}}\sin\left(k_{3_0}d\right) \right.\right.\\
        \left. \left.
        \qquad+\frac{k_{0_-}^{2}}{k_{1_0}k_{2_0}}\left(-\sin\left(k_{2_0}d\right)\cos\left(k_{3_0}d\right)+
       \frac{k_{3_0}\mu_-^2}{k_{2_0}}\sin\left(k_3d\right)\cos\left(k_{2_0}d\right)\right)\right)^2\right]^{-1}.
        \end{array}
   \end{equation}
At resonances, the GHL shifts can be written as follows
\begin{equation} \label{eq 37}
      s_{t}|_{k_{2_0}d=N_1\pi,{k_{3_0}d}=N_2\pi}=-d\tan\theta_{1_0}\left[\frac{k_{0_-}^2}{k_{2_0}^2}\left(1+\mu_-^2\right)+
     (-1)^{N_1}\frac{k_{12}^-k_{23}^-k_{y_0}^2}{k_{2_0}^2k_{3_0}^2}\right].
  \end{equation}

Having obtained the closed form expressions of the GHL shifts in different
energy domains, we proceed now to compute these quantities numerically.
This will help us understand the effect of various potential parameters on the GHL shifts
in our double barrier structure.

\section{Discussions}

\begin{figure}[h]
\centering
  \includegraphics[width=10cm, height=6cm]{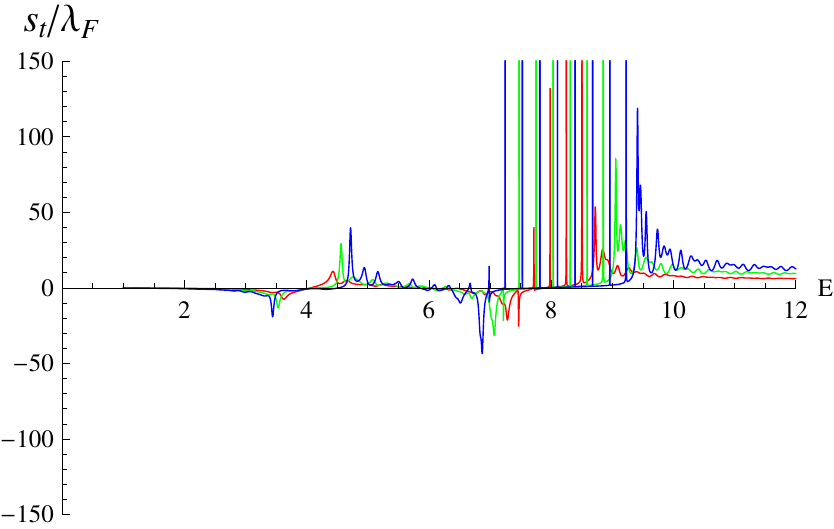}\\ 
\ \ \  \ \ \includegraphics[width=10cm, height=6cm]{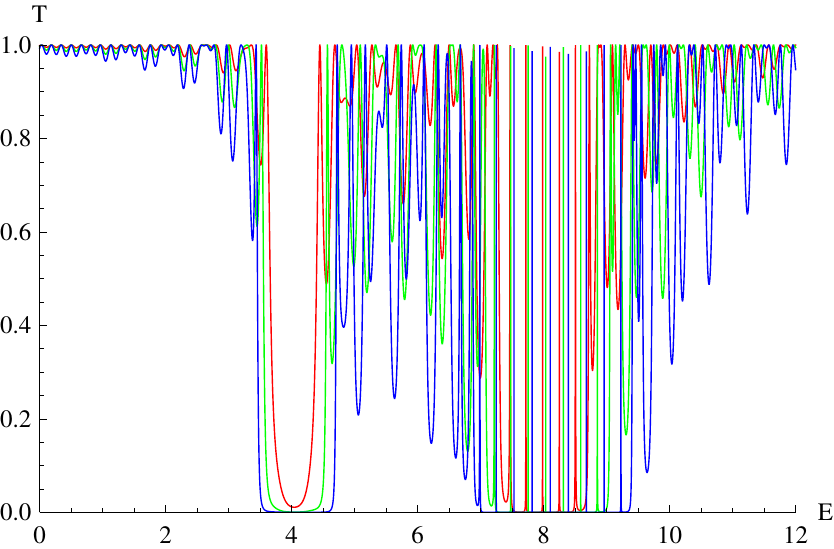}
 \caption{\sf {The GHL shifts and the transmission as a function of energy, with
  $d=10 nm$, $v=8 meV$, $u=4 meV$, $\theta_{1_0}= 4^\circ$ (red line), $\theta_{1_0}=
 6^\circ$ (green line), $\theta_{1_0}= 8^\circ$ (blue
line).}}\label{fig.ste}
\end{figure}

To allow for a suitable interpretation of our main results, we
compute numerically the GHL shifts under various conditions. First
we plot the GHL shifts as a function of the energy for
specific values of the potential parameters ($d=10 nm$, $v=8 meV$,
$u=4 meV$) and  three different values of the incidence angle
$\theta_{1_0}=4^\circ, 6^\circ, 8^\circ$, see Figure \ref{fig.ste}.
It is clear from this figure that the GHL shifts change sign at
the Dirac points, namely $(E=u,E=v)$. We deduce that there is a
strong dependence of the GHL shifts on the incidence angle
$\theta_{1_0}$, it increases with $\theta_{1_0}$. As observed in
the work of Chen \cite{Chen3}, it is shown that the GHL shifts are
related to the transmission gap $\Delta E = 2\hbar k_yv_F$. We
notice that the GHL shifts displays sharp peaks inside the
transmission gap around the point $E=v$, while they are absent
around the energy point $E=u$. {In such situation,
one can clearly end up with an interesting result such that
the number of sharp peaks is equal of that of transmission resonances.}
We also observe that the shifts become constant after certain threshold
energy value, which is compatible with a maximum of transmission.\\

It is interesting to investigate how the GH shifts behave as a
function of the barrier potential heights, i.e. $v$ and $u$, the
numerical results are shown in Figure 3. We  have chosen the
parameters ($E=4$, $v=8$) in Figure 3(a) and ($E=8$, $u=4$) in
Figure 3(b), with inter-barrier distance $d=10nm$ and angles
$\theta_{1_0}=4^\circ,6^\circ,8^\circ$. One can notice that, at
the Dirac points ($E=u$,$E=v$), the GH shifts change their sign
and behave differently as compared to Figure 2. This change in
sign of the GH shifts shows clearly that they are strongly
dependent on the barrier heights. We also notice that the GH
shifts are positive as long as the energy satisfies the condition
$E>v>u$ and negative for $E<u<v$. However, in the energy domain
$u<E<v$ Figures 3(a) and 3(b) show different behaviors such that
the shifts are negative and positive, respectively. Note that, the
Dirac points represent the zero modes for Dirac operator
\cite{Sharma} and lead to the emergence of new Dirac points, this
point has been discussed in different works
\cite{Bhattacharjee,Park1,Park2}. Such point separates the two
regions of positive and negative refraction. In the cases of $v<E$
and $v>E$ (respectively $u<E$ and $u>E$), the shifts are
respectively in the forward and backward directions, due to the
fact that the signs of group velocity are opposite.

\begin{figure}[h]
\label{fig.stuv}
\centering
\includegraphics[width=8cm, height=6cm]{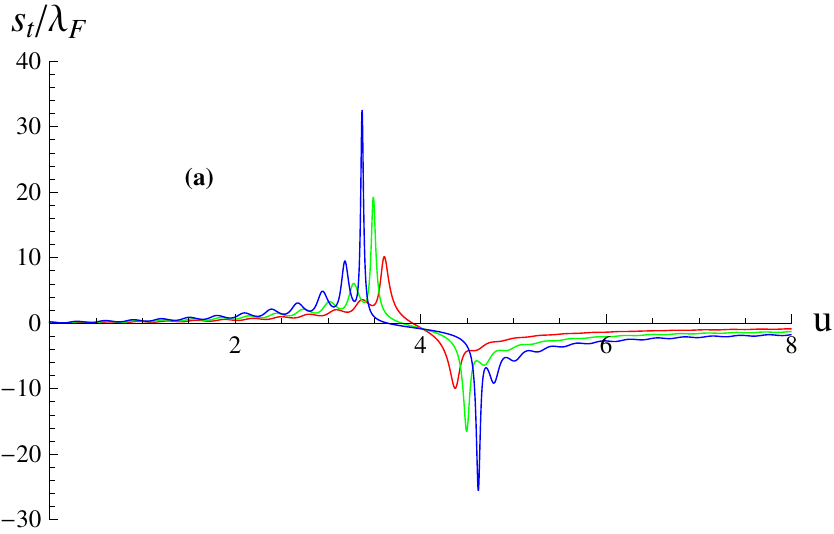}\ \ \ \
\includegraphics[width=8cm, height=6cm]{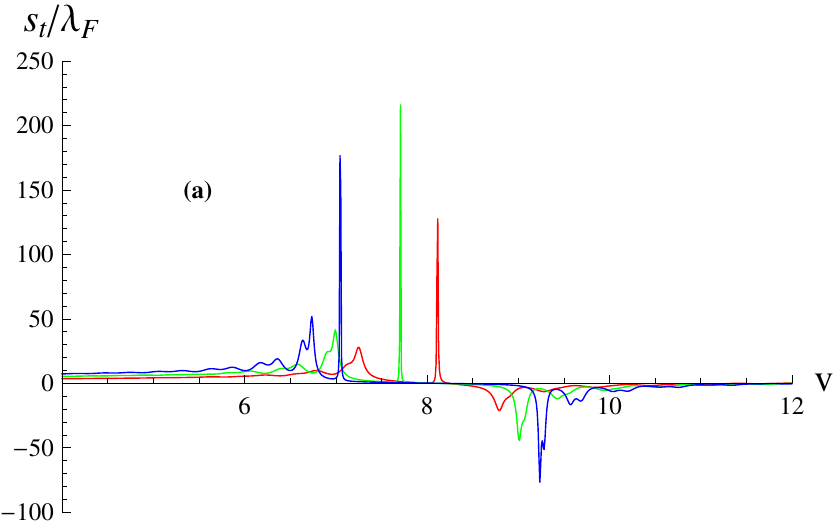}\\
\ \ \ \includegraphics[width=8cm, height=6cm]{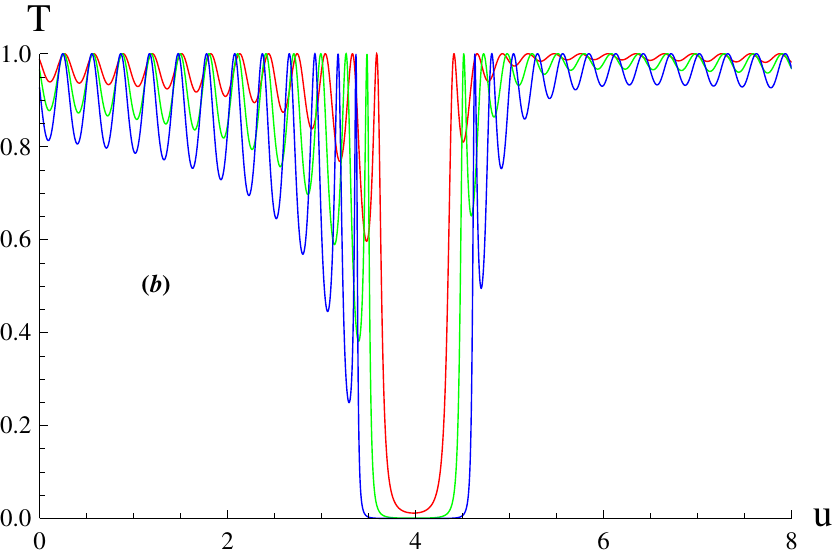}\ \ \ \ \
\includegraphics[width=8cm, height=6cm]{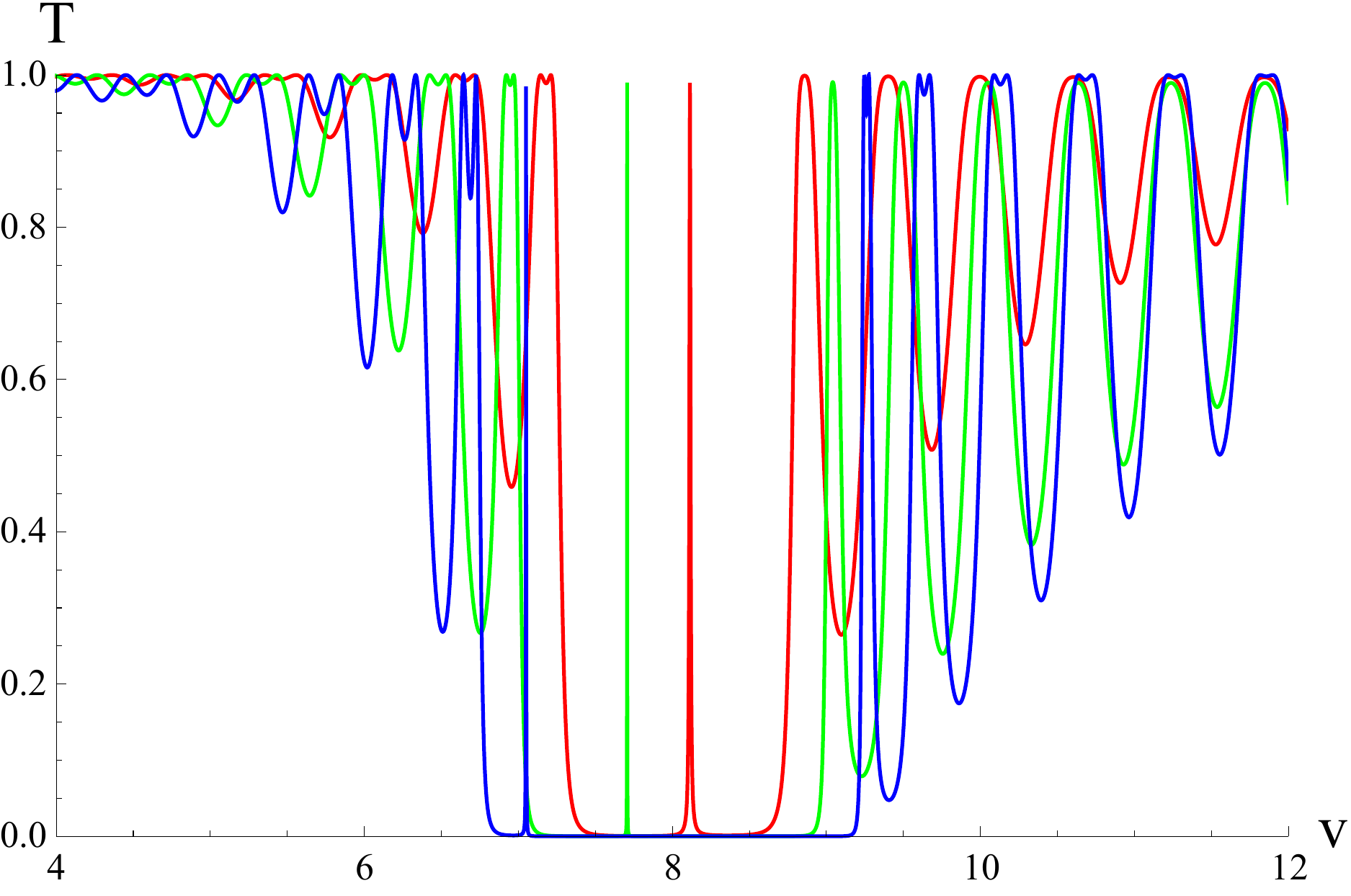}
  \caption{\sf {The GHL shifts and the transmission as a function of the heights $u$ and $v$ of the potential barrier. (a)/(b)
  for height $u/v$, with ($E=4$, $v=8$)/($E=8$, $u=4$), where $d=10 \,nm$, $\theta_{1_0}= 4^\circ$ (red line), $\theta_{1_0}= 6^\circ$ (green line),
  $\theta_{1_0}= 8^\circ$ (blue line).}}
\end{figure}

\begin{figure}[h]
  \centering
  \includegraphics[width=10cm, height=6cm]{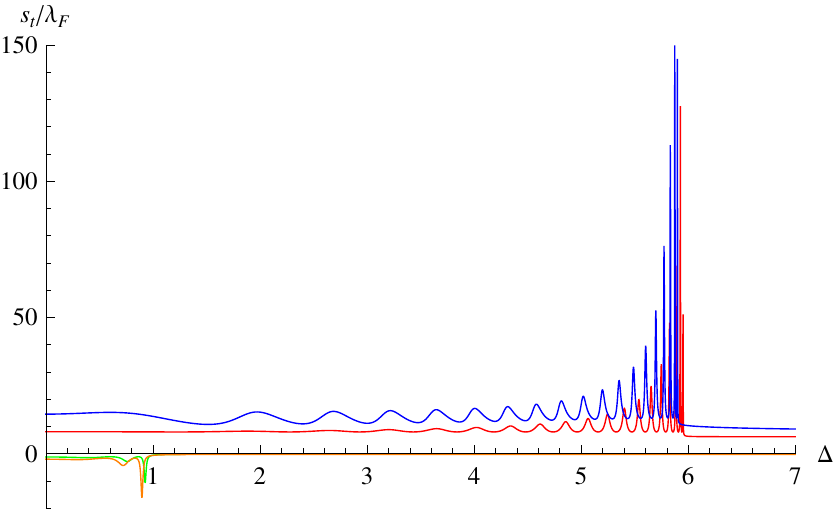} \\ 
  \ \ \includegraphics[width=10cm, height=6cm]{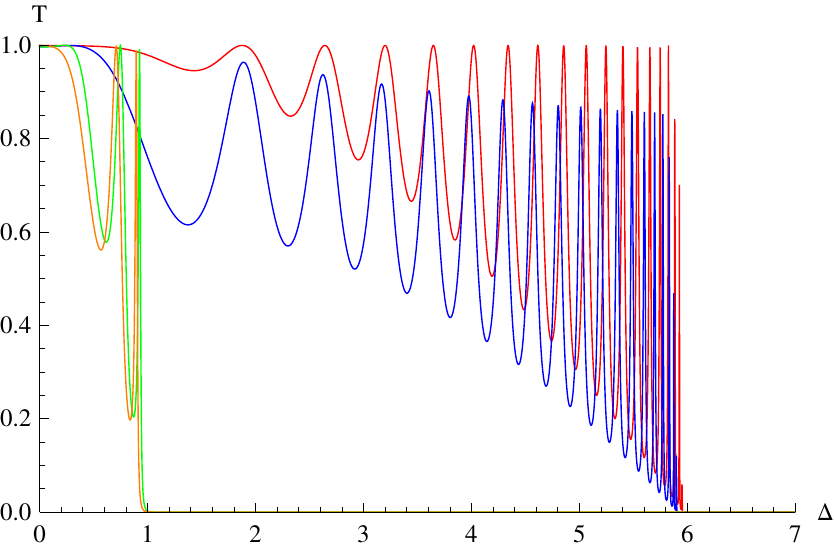}
  \caption{\sf {The influence of the induced gap $\Delta$ on the GHL shifts and the transmission
  in the presence of a double barrier, for $d=10 nm$, $v=8 meV$, $u=4meV$, $E=10 meV$, $\theta_{1_0}= 4^\circ$ (red line), $E=10 meV$,
$\theta_{1_0}= 6^\circ$ (blue line), $E=3 meV$, $\theta_{1_0}=
4^\circ$ (green line), $E=3 meV$, $\theta_{1_0}= 6^\circ$ (orange
line).}}\label{st.delta}
\end{figure}

Now let us investigate what will happen if we introduce a gap in
the band structure. Note that, the gap is introduced as shown in
Figure 1 and therefore it affects the system energy according to
the solution of the energy spectrum obtained in region 3. Figure
\ref{st.delta} shows that the GHL shifts in the propagating case
can be enhanced by a gap opening at the Dirac point. This has been
performed by fixing the parameters $d=10nm$, $v=8meV$, $u=4meV$
and making different choices for the energy and angle. For the
configuration ($E=3meV$, $\theta_{1_0}=6^\circ$) we conclude that
we can still have negative shifts as it is shown by the orange
line. However for other configurations, we do not have such
behavior, more specifically by increasing $\Delta$, the GHL shifts
become mostly constant up to some value then show sharp peaks as
indicated in blue color where we have taken $E=10meV$ and $\theta_{1_0}=6^\circ$.  \\

\begin{figure}[h]
  \centering
  \includegraphics[width=10cm, height=5cm]{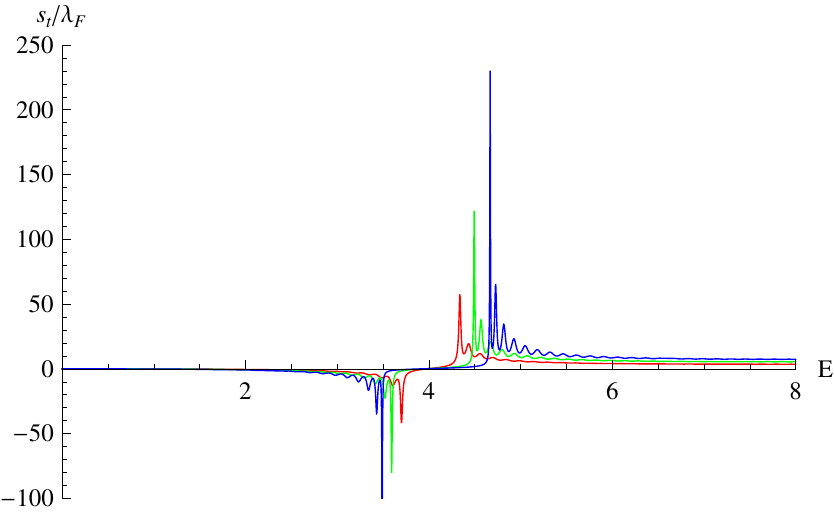}\\ 
\ \ \ \    \ \includegraphics[width=10cm, height=5cm]{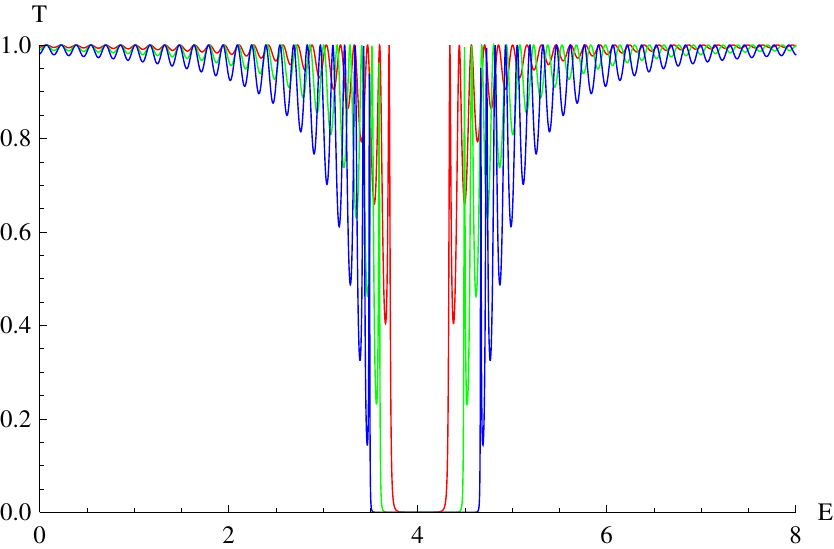}
  \caption{\sf{The GHL shifts and the transmission as function of the energy $E$,
  for $d=10 nm$, $v=u=4 meV$, $\theta_{1_0}= 4^\circ$ (red line), $\theta_{1_0}=
  6^\circ$ (green line), $\theta_{1_0}= 8^\circ$ (blue line). }}\label{fig.stu=v}
\end{figure}

To make comparison with the relevant literature and show the importance of our results, we
will discuss three interesting special cases.
Since our work is a generalization of \cite{Chen4} to double barriers,
we first show how to recover their results. This can be done by requiring that
$u = v$, which implies that
 $k_{F_2}=k_{F_3}$ or $k_{2_0}=k_{3_0}$ and therefore the present system
behaves like a single barrier where the GHL shift reduces to
 \begin{equation}\label{st_u=v}
    s_{t}=2dT\tan\theta_{1_0}\left[\left(2\pm \left(\frac{k_{0\pm}^2}{k_{1_0}^2}+\frac{k_{0\pm}^2}{k_{2_0}^2}\right)\right)
\frac{\sin\left(4k_{2_0}d\right)}{4k_{2_0}d}\mp\frac{k_{0\pm}^2}{k_{2_0}^2}\right]
 \end{equation}
and the corresponding transmission coefficient is given by
 \begin{equation}\label{trans_u=v}
    T=\left[\cos^2\left(2k_{2_0}d\right)+\frac{k_{0\pm}^4}{k_{2_0}^2k_{1_0}^2}\sin^2\left(2k_{2_0}d\right)\right]^{-1}
 \end{equation}
where $\pm$ correspond to Klein tunneling and classical behavior,
respectively. Note that these results are identical to those obtained previously in
\cite{Chen4}.
The above GHL shifts and transmission are plotted in  Figure \ref{fig.stu=v}. It is clearly
seen that $s_t$ is oscillating between negative and positive values around the critical point
$E=u=v$. At such a point $T$ is showing zero transmission while it oscillates away from the critical point.\\

\begin{figure}[h]
  \centering
  \includegraphics[width=10cm, height=6cm]{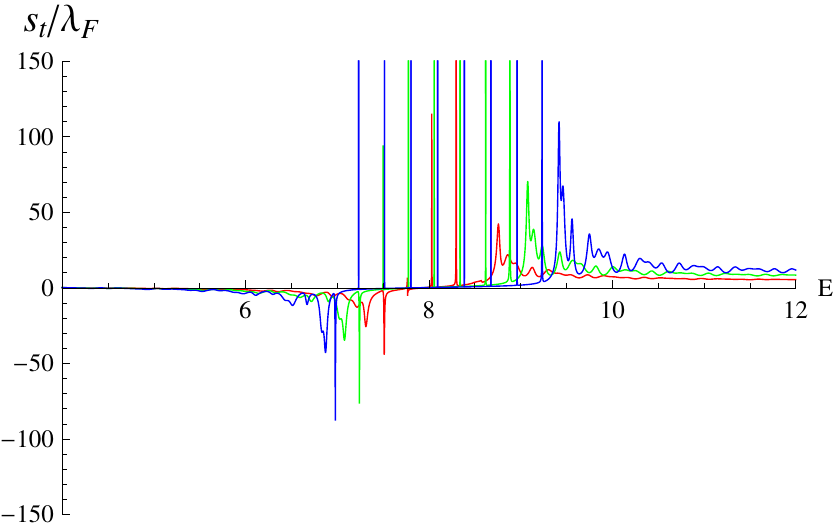} \\ 
  \ \ \ \  \includegraphics[width=10cm, height=6cm]{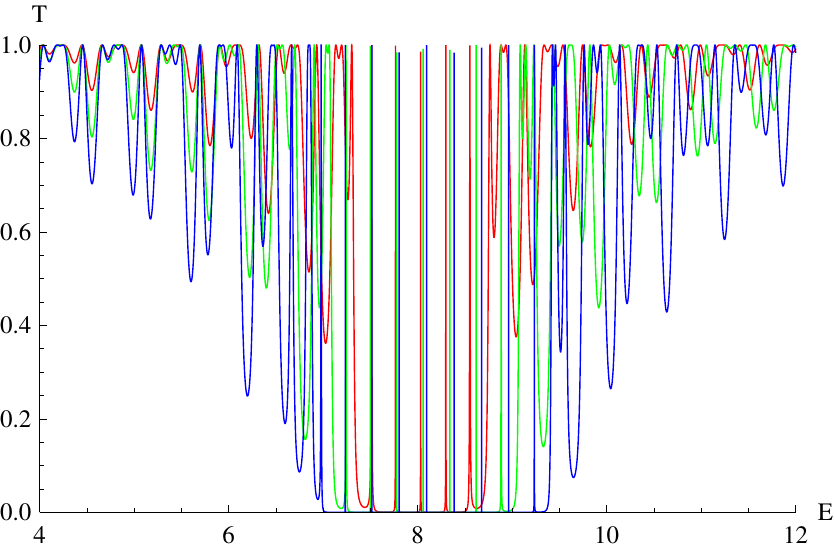}
  \caption{\sf{The GHL shifts and the transmission as function of the energy $E$,
  with $d=10nm$, $v=8meV$, $u=0meV$, $\theta_{1_0}=4^\circ$ (red line), $\theta_{1_0}=6^\circ$
  (green line), $\theta_{1_0}=8^\circ$ (blue line). }}\label{fig.stu=0}
\end{figure}

In the second particular case we consider $u=0$, which implies
equality between wave vectors $k_{F_1}=k_{F_3}$, and is equivalent
to the requirement that $k_{1_0}=k_{3_0}$. This is similar to one
potential configuration for the system studied in \cite{Song} to
deal with the GHL shifts. In the present  case, $s_t$ becomes
\begin{equation}\label{eq }
        s_{t}=-d^2T\tan{\theta_{1_0}}\left(-A_{\pm}^{'}B_{\pm}^{'}+C_{\pm}^{'}D_{\pm}^{'}\right)
 \end{equation}
where we have set
\begin{equation} \label{eq 27}
    \begin{array}{llll}
           A_{\pm}^{'}=\frac{1\mp\mu_{\pm}^2}{k_{2_0}d}\cos\left(k_{3_0}d\right)\sin\left(k_{2_0}d\right)+
           \frac{1\mp\nu_{\pm}^2}{k_{3_0}d}\cos\left(k_{2_0}d\right)\sin\left(k_{3_0}d\right)+
           \frac{2\pm\mu_{\pm}^2 \pm\nu_{\pm}^2}{k_{2_0}k_{3_0}d^2}\sin\left(k_{2_0}d\right)
           \sin\left(k_{3_0}d\right)\\
           B_{\pm}^{'}=\frac{k_{23}^{2\pm}k_{y_0}^2}{k_{2_0}^2k_{3_0}}\sin\left(k_{3_0}d\right)\pm
           k_{2_0}\nu_{\pm}^2\left(\cos\left(k_{3_0}d\right)\sin\left(k_{2_0}d\right)\mp
           \frac{k_{3_0}\mu_{\pm}^2}{k_{2_0}}\sin\left(k_{3_0} d\right)\cos\left(k_{2_0}d\right)\right)\\
           C_{\pm}^{'}=\cos\left(k_{2_0}d\right)\cos\left(k_{3_0}d\right)\pm\frac{k_{3_0}\mu_{\pm}^2}{k_{2_0}}\sin\left(k_{2_0}d\right)\sin\left(k_{3_0}d\right)\\
           {D_{\pm}^{'}={\mp}k_{2_0}\nu_{\pm}^2\left(\frac{1\mp\mu_{\pm}^2}{k_{2_0}d}\cos\left(k_{2_0}d\right)\cos\left(k_{3_0}d\right)-
           \frac{1\mp\nu_{\pm}^2}{k_{3_0}d}\sin\left(k_{2_0}d\right)\sin\left(k_{3_0}d\right)+
           \frac{2\pm\mu_{\pm}^2\pm
          \nu_{\pm}^2}{k_{2_0}k_{3_0}d^2}\cos\left(k_{2_0}d\right)\sin\left(k_{3_0}d\right)\right)}\\{\qquad+
           \frac{k_{23}^{2\pm}}{k_{2_0}^2k_{3_0}d^2}\left(2\sin{k_{3_0}d}+
           2k_{y_0}^2\left(\frac{1}{k_{2_0}^2}+\frac{1}{k_{3_0}^2}\right)\sin{\left(k_{3_0}d\right)}-
           \frac{k_{y_0}^2d}{k_{3_0}}\cos\left(k_{3_0}d\right)\right)}\\{\qquad+\frac{1}{k_{2_0}d^2}
            \left(2\pm\mu_{\pm}^2\pm\nu_{\pm}^2\right)\left(
           \cos\left(k_{3_0}d\right)\sin\left(k_{2_0}d\right)\mp\frac{k_{3_0}\mu_{\pm}^2}{k_{2_0}}\cos\left(k_{2_0}d\right)
           \sin\left(k_{3_0}d\right)\right)}.\nonumber
       \end{array}
 \end{equation}
Transmission probability is given by
\begin{equation} \label{eq 28}
    \begin{array}{ll}
    T =\left[\left(\cos\left(k_{2_0}d\right)\cos\left(k_{3_0}d\right)\pm
         \frac{k_{3_0}\mu_{\pm}^2}{k_{2_0}}\sin\left(k_{2_0}d\right)\sin\left(k_{3_0}d\right)\right)^2
          +\left(\frac{k_{23}^{2\pm}k_{y_0}^2}{k_{2_0}^2k_{3_0}^2}\sin\left(k_{3_0}d\right)
           \right.\right.\\
          \left. \left. \qquad \pm\mu_{\pm}\nu_{\pm}\left(\sin\left(k_{2_0}d\right)\cos\left(k_{3_0}d\right)-
          \frac{k_{3_0}\mu_{\pm}^2}{k_{2_0}}\sin\left(k_{3_0}d\right)\cos\left(k_{2_0}d\right)\right)\right)^2\right]^{-1}
    \end{array}
  \end{equation}
where the sign $\pm$ corresponds to the two energy intervals,
$0<E<v$ $(-)$ and $E>v$ $(+)$, respectively. The GHL shifts and
transmission as functions of the energy $E$ are shown in Figure
\ref{fig.stu=0} for the values $d=10nm$, $v=8meV$, $u=0meV$  and
$\theta_{1_0}=4^\circ, 6^\circ, 8^\circ$. Both quantities are
showing a series of peaks and resonances. The resonances
correspond to the bound states of the double barriers. We notice
that the GHL shifts peak at each bound state energy and are
clearly shown in the transmission curve underneath. The energies
at which transmission vanishes correspond to energies at which the
GHL shifts change sign. Since these resonances are very sharp
(true bound states with zero width) it is numerically very
difficult to track all of them, if we do then the alternation in
sign of the GHL shifts will be observed.
{As before, we notice that around the Dirac point $E=v$ the number
of peaks is equal of that of transmission resonances}.
{To summarize, we notice that a superposition of the two Figures \ref{fig.stu=v} and
\ref{fig.stu=0}, obtained in both particular cases for $u=v$ and $u=0$, respectively,
gives exactly Figure \ref{fig.ste}.}\\

\begin{figure}[h]
  \centering
  \includegraphics[width=10cm, height=6cm]{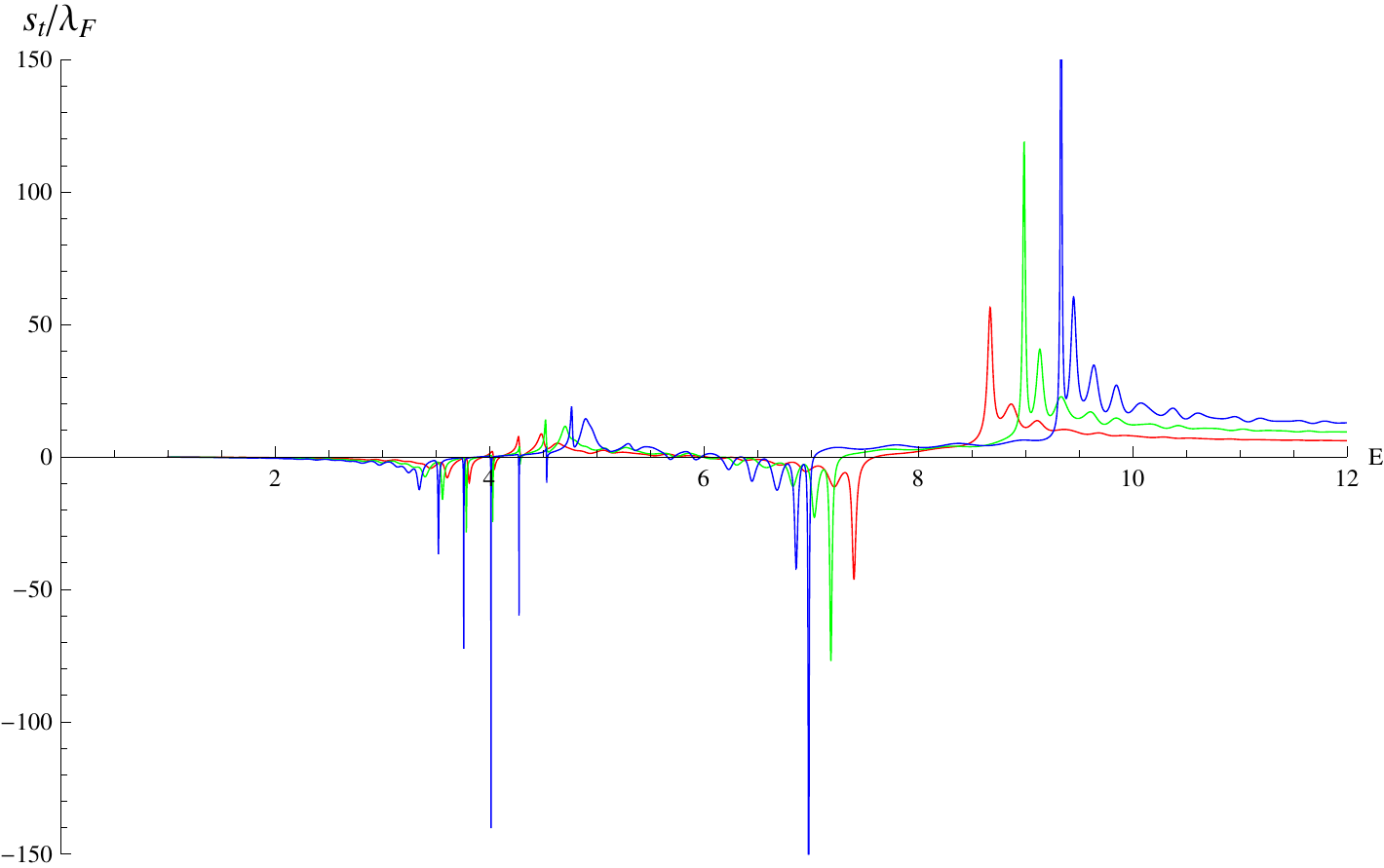}\\ 
\ \   \includegraphics[width=10cm, height=6cm]{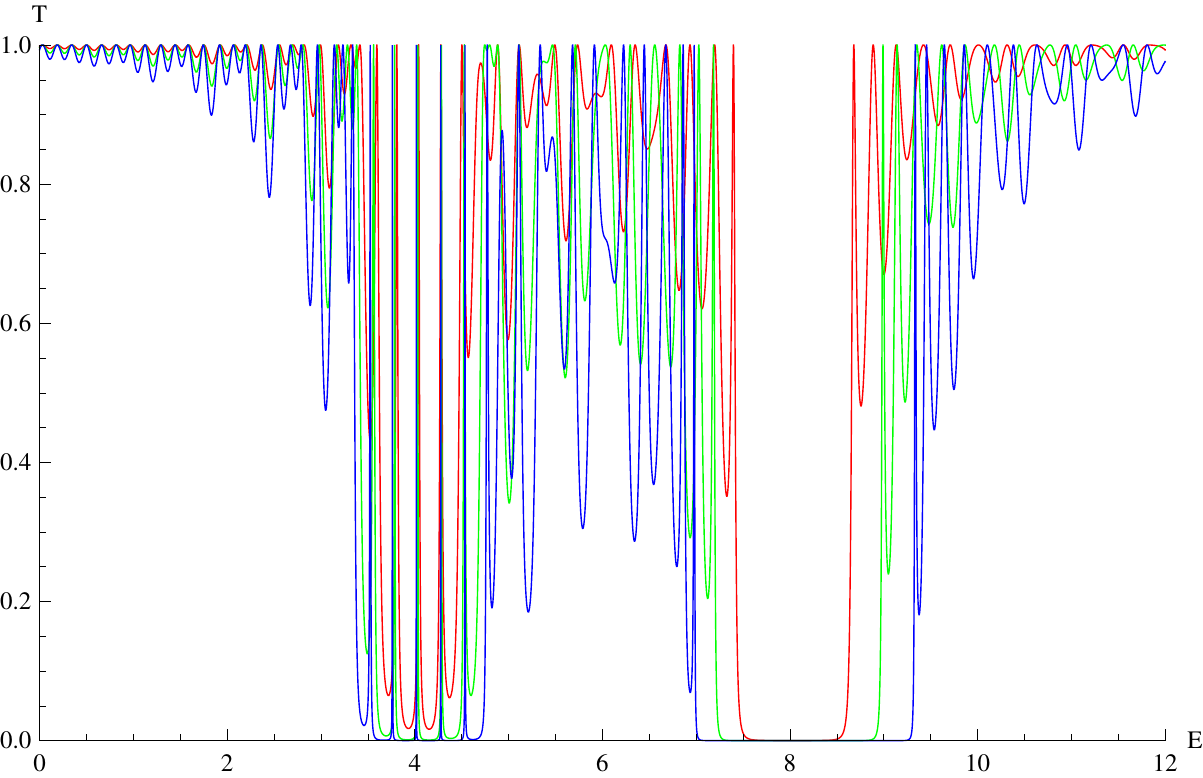}
  \caption{\sf {The GHL shifts and the transmission as a function of the energy $E$ in the case where $u>v$, with $d=10nm$, $v=4meV$, $u=8meV$,
  $\theta_{1_0}=4^\circ$ (red line),
  $\theta_{1_0}=6^\circ$ (green line), $\theta_{1_0}=8^\circ$ (blue line).}}\label{fig.st_u>v}
\end{figure}

Finally we consider the third case  where $u>v$. In such situation the GHL shifts and transmission are
displayed in Figure \ref{fig.st_u>v}.
{It is clearly seen that
the GHL shifts display sharp peaks inside the transmission gap around the point the Dirac point $E=v$ but no peaks  inside the transmission gap around the point Dirac point $E=u$. This happened in the negative region rather than positive one as seen in Figure \ref{fig.ste}, which is due to the exchange of role between barrier heights $u$ and $v$, i.e. $u>v$. Due to this exchange the structure of the transmission is reversed between Figure 2 and Figure 7, sharp resonances appear in the higher transmission gap in Figure 2 while they appear in the lower transmission gap in Figure 7. As observed in previous figures, we have the same numbers of sharp peaks in $s_t$ and transmission resonances. Thus in our situation we have transmission gaps and zero-k gaps which are identified with Dirac points, in our case with energies $E = v$ and $E = u$. We have observed that GHL shifts change sign when crossing the edges of the transmission gap but is not affected by the zero-$k$ gaps. This observation was not affected by the incident angle as long as it is below the critical angle defining total reflection. It is worth mentioning that these observations regarding zero-k gap are in contrast to those in superlattices where GHL shifts change sign at the edge of the zero-$k$ band gap. }

In summary, we note that the GHL shifts display sharp peaks inside the transmission gap around the point Dirac $E=v$ in our system as shown in Figure \ref{fig.1}. These peaks can be attributed to the quasibound states formed in the double barrier structure. To confirm these findings we have also studied the simple barrier structure as a special case of our potential configuration by setting $u=v$, these peaks are then absent. The GHL shifts inside the transmission gap around the point $E=u$ in our system is the same as in the simple barrier case \cite{Chen4}.

\newpage
\section{Conclusion}

We have computed the Goos-H\"anchen like (GHL) shifts through a double barrier potential in a single layer graphene system. The massless Dirac-like equation was used to describe the scattered fermions by such potential configuration. Our results show that the GHL shifts is affected by the internal structure of the double barrier, in particular the GHL shifts change sign at the transmission zero energies and peaks at each bound state associated with the double barrier. Thus our numerical results show that the GHL shifts can be enhanced by the presence of resonant energies in the system when the incident angle is less than the critical angle associated with total reflection.

It was also observed that the transmission gap increases with the
incidence angle as long as it less than the critical angle. The
gap within the well region is seen to reduce both transmission and
GHL shifts which exhibit an oscillatory behavior as a function of
the energy gap. The GHL shifts also depend on the potential
parameters, more specifically the heights of the barrier and well
regions, $u$ and $v$. In particular for $v > u$ we observe that
there is no Klein region while for $u > v$ we do have a Klein
tunneling region which enhances transmission and GHL shifts. Thus
with double barrier structure we can have more control on the GHL
shifts.
To support the validity of our findings we have selected our potential parameters so as to reduce it to a single
barrier and confirmed all results found previously by other groups \cite{Chen4}. In the case of a single barrier peaks corresponding to bound states are absent in the zero transmission region. Also we have checked the results obtained in \cite{Song} for the particular case $u=0$.

{Finally, we close our work by mentioning some challenges facing
the potential connection between two fields: quantum optics and graphene. Very recently,
pertinent discussions have been made to emphasis the main difficulties in detecting the Goos-Hanchen shifts
and preparing the electron beam in solid-state physics \cite{chenrev}. These discussions open for us important
research avenues that will help us understand and overcome the above mentioned difficulties. On the other hand,
we learned from \cite{chenrev} that the spin-orbit coupling in optics is an interesting and fascinating
topic because the spin-orbit interaction in graphene opens up a spin-orbit gap, though very small, at the Dirac points. All these matters will be highly important when we consider tunable GHL shift leading to potential applications in future graphene based electronic devices. These matters will be investigated in the near future to enable us to get a deeper understanding of graphene transport properties. }

\section*{Acknowledgments}
The authors would like to acknowledge the support of King Fahd University of Petroleum and minerals
under the theoretical physics research group project RG1306-1 and RG1306-2. The generous support
provided by the Saudi Center for Theoretical Physics (SCTP) is highly appreciated by all authors.
AJ  thanks the Deanship of Scientific Research at King Faisal University for funding this research number  (140232).

\end{document}